\documentclass[modern]{aastex62}

\shorttitle{}
\shortauthors{}

\begin{document}

\title{Comment on ``An X-ray chimney extending hundreds of parsecs above and below the Galactic Centre'' (2019, Nature, 567, 34)}

\email{shinya.nakashima@riken.jp}

\author{Shinya Nakashima}
\affil{RIKEN High Energy Astrophysics Laboratory, 2-1 Hirosawa, Wako, Saitama, 351-0198, Japan}

\author{Katsuji Koyama}
\affil{Department of Physics, Graduate school of Science, Kyoto University, Kitashirakawa-Oiwake-cho, Kyoto, 606-8502, Japan}

\author{Q. Daniel Wang}
\affil{Department of Astronomy, University of Massachusetts, Amherst, MA 01003, USA}

\section*{} 
A recent article ``An X-ray chimney extending hundreds of parsecs above and below the Galactic Centre" (2019, Nature, 567, 34) reported the detection of chimney-like X-ray-emitting features  above and below the Galactic Center from XMM-Newton observations. We note here that these features were already reported by our Suzaku papers:  Nakashima et al. (2013, ApJ, 773, 20, arXiv:1310.4236) for the southern feature and Nakashima et al. (2019, ApJ, in press, arXiv:1903.02571) for the northern feature.
In particular, Nakashima et al. (2013) show that the ionization state of the southern feature is not in collisional ionization equilibrium and is most likely in a recombining or over-ionized state, which suggests its origin in the Galactic Center about 0.1 Myr ago.



\end{document}